\documentstyle[prl,aps,twocolumn,epsfig]{revtex}
\begin{document}


\title{Single nanoparticle measurement techniques}

\author{W. Wernsdorfer$^1$, D.~Mailly$^2$, A. Benoit$^3$}
\address{
$^1$Lab. L. N\'eel, CNRS, associ\'e \`a l'Universit\'e J. Fourier, 
BP 166, 38042 Grenoble Cedex 9, France\\
$^2$Lab. de Microstructures et de Micro\'electronique, 196 av. H. Ravera, 92220 
Bagneux, France\\
$^3$CRTBT, CNRS, associ\'e \`a l'Universit\'e J. Fourier, 
BP 166, 38042 Grenoble Cedex 9, France
}

\date{\today~~{\bf MMM~1999~BE--01}}
\maketitle
\begin{abstract}
{\bf Various single particle measuring techniques are briefly reviewed and the basic concepts 
of a new micro-SQUID technique are discussed. It allows measurements of 
the magnetization reversal of single nanometer-sized 
particles at low temperature. 
The influence of the measuring technique on the system of interest is discussed.}
\end{abstract}

\narrowtext

The dream of measuring the magnetization reversal of an individual magnetic particle goes 
back to the pioneering work of N\'eel \cite{Neel49}. The first realization was published by 
Morrish and Yu in 1956 \cite{Morrish56}. 
These authors employed a quartz--fiber torsion 
balance to make magnetic measurements on individual micrometer sized 
$\gamma -$Fe$_2$O$_3$ particles. With their technique, they wanted to avoid the complication of 
particle assemblies which are due to different orientations of the particle's easy axis of 
magnetization and particle--particle dipolar interaction. 
They attempted to show the existence 
of a single--domain state in a magnetic particle. Later on, other groups tried to study single 
particles but the experimental precision did not allow a detailed study. A first 
breakthrough came via the work of Knowles \cite{Knowles78} who developed a simple 
optical method for measuring the switching field, defined as the minimum applied field 
required to reverse the magnetization of a particle. However, the work of Knowles failed to 
provide quantitative information on well defined particles. 

More recently, insights into the 
magnetic properties of individual and isolated particles were obtained with the help of 
electron holography \cite{Tonomura86}, vibrating reed magnetometry \cite{Richter89}, 
Lorentz microscopy \cite{Heffermann91,Salling91}, and magnetic force microscopy 
\cite{Chang94,Ledermann94}. Most of the studies have been carried out using 
magnetic force microscopy at room temperature. This technique has an excellent spatial 
resolution but dynamical measurements are difficult due to the sample--tip interaction 
\cite{Rave94}. Only a few groups could study the magnetization reversal of individual 
nanoparticles or nanowires at low temperatures. 

The first magnetization measurements of 
individual single-domain nanoparticles and nanowires at very low temperatures were 
presented by Wernsdorfer {\it et al.} \cite{Wernsdorfer95}. 
The detector, a Nb micro-bridge-Superconducting 
Quantum Interference Device (SQUID), and the studied particles were 
fabricated using electron-beam lithography. By measuring the electrical resistance of 
isolated Ni wires with diameters 
between 20 and 40 nm, Giordano and Hong studied the motion of magnetic domain walls 
\cite{Hong95,Wegrowe98}. Other low temperature techniques which may be adapted to 
single particle measurements are Hall probe 
magnetometry \cite{Kent94,Geim98}, magnetometry 
based on the giant magnetoresistance \cite{Cros97,Gallagher97} or spin-dependent 
tunneling with Coulomb blockade \cite{Doudin97,Gueron99}. At the time of writing, the 
micro-SQUID technique allowed the most detailed study of the magnetization reversal of 
nanometer-sized particles \cite{Wernsdorfer_PRLs}. In the 
following, we review the basic ideas of the micro-SQUID technique.

\begin{figure}[b]
\centerline{\epsfxsize=6.5 cm \epsfbox{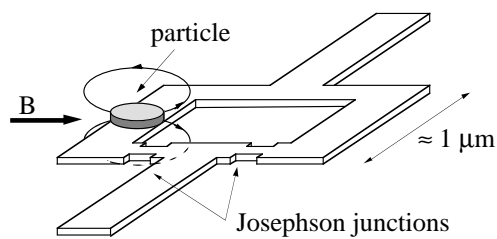}}
\caption{Position of the particle on the SQUID loop.}
\label{SQUID}
\end{figure}

The Superconducting Quantum Interference Device (SQUID) has been used very 
successfully for magnetometry and voltage or current measurements in the fields of 
medicine, metrology and science. SQUIDs are mostly 
fabricated from a Nb$-$AlO$_x-$Nb trilayer, several hundreds of nanometers thick. The 
two Josephson-Junctions are planar tunnel junctions with an area of micrometer size. In 
order to avoid flux pinning in the superconducting film, the SQUID is placed in a 
magnetically shielded environment. The sampleÕs flux is transferred via a superconducting 
pick--up coil to the input coil of the SQUID. 
Such a device is widely used as the signal can 
be measured by simple lock--in techniques. 
However, this kind of SQUID is not well 
suited to measuring the magnetization of single submicron-sized samples as the separation 
of SQUID and pickup coil leads to a relatively small coupling factor. A much better 
coupling factor can be achieved by coupling the sample directly with the SQUID loop. In 
this arrangement, the main difficulty arises from the fact that the magnetic field applied to 
the sample is also applied to the SQUID. 

Our choice of the micro-SQUID configuration was mainly motivated by the simplicity of 
fabrication, by the desired temperature range, and by the lack of sensitivity to a high field 
applied in the SQUID plane. These criteria led to the use of micro bridge junctions instead 
of the commonly used tunnel junctions. 

\begin{figure}[b]
\centerline{\epsfxsize=6.5 cm \epsfbox{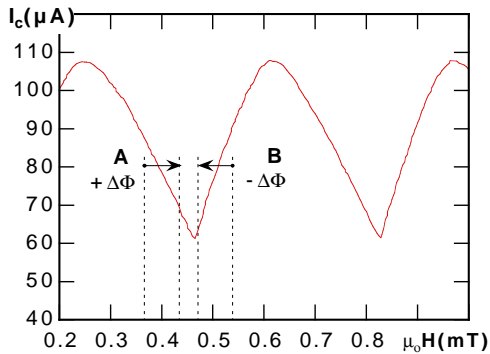}}
\caption{Critical current Ic of a micro-SQUID as a function of a magnetic field applied 
perpendicular to the SQUID plan. For cold mode method , SQUID is biased close to the 
critical current so that the it is in state A or B, respectively of positive or negative flux 
jumps induced by the magnetization reversal.}
\label{fig_caract}
\end{figure}

The Josephson effect in microbridges has first been suggested in 1964 by Anderson and 
Dayem \cite{Anderson64}. These superconducting weak links seemed to be very 
promising in order to design planar DC-SQUIDs with a one step thin film technology. 
However, Dayem bridges exhibit a Josephson current-phase relation only when their 
dimensions are small compared to the coherence length $\xi$. 
Nowadays, electron beam lithography allows one to directly fabricate reliable microbridge 
Josephson Junctions made of materials like Al, Nb and Pb. Benoit {\it et al.} developed 
micrometer-sized SQUIDs having two microbridge Josephson Junctions 
(Fig.~\ref{SQUID})\cite{Mailly93}. 
These SQUIDs have a hysteretic $I-V$ curve which made it 
impossible to use standard SQUID electronics to read out the SQUID. Therefore, their 
method consists in measuring the critical current of the SQUID loop. As the critical 
current $I_c$ is a periodic function of the flux going through the SQUID loop (cf. 
Fig.~\ref{fig_caract}), one can easily deduce the flux change in the SQUID loop by 
measuring the critical current.

By using electron beam lithography, planar Nb micro-bridge-DC-SQUIDs (of 1 to 2 
$\mu$m in diameter) were constructed on which a magnetic 
particle was placed (Fig.~\ref{SQUID}) \cite{WW_PhD}. 
The SQUID detected the flux through its loop produced by the 
sample's magnetization. The sensitivity achieved by the critical current measurement 
technique was about $10^{-4} \Phi_0/\sqrt{Hz}$ 
($\Phi_0$ = h/2e = $2 \times 10^{-15}$ Wb). For hysteresis 
loop measurements, the external field was applied in the plane of 
the SQUID, thus the SQUID is only sensitive to the flux induced by the stray field of the 
sampleÕs magnetization. Due to the proximity between sample and SQUID, 
magnetization reversals corresponding to $10^{3} \mu_B/\sqrt{Hz} (10^{-17} 
emu/\sqrt{Hz})$ could be detected, i.e. the magnetic moment of a Co nanoparticle with a 
diameter of 2-3 nm.

In order to have a SQUID which can be exposed to a high field applied in the SQUID 
plane, the SQUID was made out of a very thin layer preventing flux trapping. For the 
experimental set-up, we used a 10 - 20 nm thick Nb layers allowing measurements of 
hysteresis loops in magnetic fields of up to 2 T and at temperatures below 7 K. The time 
resolution was given by the time between two measurements of the critical current. In this 
case, the achieved time resolution was 100 $\mu$s.

In order to study the temperature dependence of the magnetization reversal and 
macroscopic quantum tunneling of magnetization, the main difficulty associated with the 
SQUID detection technique lies in the Joule heating when the critical current is reached. 
After the normal-state transition at the critical current, the SQUID dissipates for about 100 
ns which slightly heats the magnetic particle coupled to the SQUID. This problem can be 
solved by using the SQUID only as a trigger \cite{Wernsdorfer_PRLs}. 

In the superconducting state, the SQUID is biased close to the critical current and a field 
is applied perpendicular to the SQUID plane so that the SQUID is in state A or B of 
Fig.~\ref{fig_caract}, respectively of the positive or negative flux jumps induced by the 
magnetization reversal. The magnetization reversal of the particle then triggers the 
transition of the SQUID to the normal state. By this method, the sample is only heated 
after the magnetization reversal. We call this measuring technique the {\it cold mode 
method}.

In addition, as the SQUID is in the superconducting state before the magnetization 
reversal, the sample does not interfere with the rf-noise which is induced in oxide layer 
Josephson junctions. 

\begin{figure}[b]
\centerline{\epsfxsize=6.5 cm \epsfbox{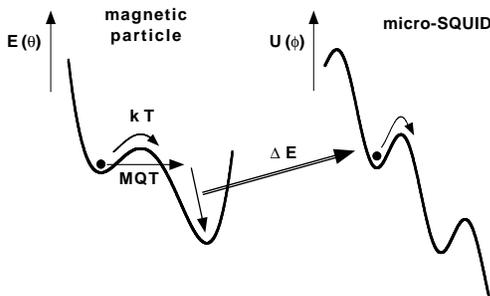}}
\caption{Energy scheme of the cold mode method. After the particle jumps over the 
saddle point or tunnels through the energy barrier, it falls into the stable potential well 
releasing energy. A very small fraction $\Delta E$ of this energy is transferred to the SQUID and 
kicks the SQUID out of its metastable superconducting state.}
\label{cold_mode}
\end{figure}

Finally, the cold mode method is very important for studying macroscopic quantum 
tunneling of magnetization. Quantum theory requires 
that the escape rate from a metastable 
potential well by quantum tunneling is strongly reduced the coupling of the magnetic 
system with its environment. Therefore, the measuring device must be weakly coupled to 
the magnetic particle. However, in order to measure the magnetization reversal, the SQUID 
must be strongly coupled to the magnetic particle which hinders the possibility of 
quantum tunneling. This problem can be solved by using the cold mode method. In order 
to show this schematically, Fig.~\ref{cold_mode} 
represents two energy potentials, one is the 
double well potential of the particle, the other is the periodic potential of the SQUID. 
Before the magnetization reversal, both systems are in a metastable state: the particle 
because of an applied field which is close to the switching field and the SQUID because 
of a current through the SQUID loop which is close to the critical current. Jumping of the 
particle over the saddle point or tunneling through the energy barrier corresponds to a 
rotation of magnetization of only a few degrees. For this starting process, the coupling 
between particle and SQUID is arranged to be very small. Afterwards, the particle falls 
into the stable well which means a rotation of magnetization of up to nearly 
180$^{\circ}$. During this process, the coupling between particle and SQUID is strong 
enough to 'kick' the SQUID out of its metastable state. The corresponding transition from 
the superconducting into the normal state is easily detected for a hysteretic SQUID.

In order to illustrates these couplings, let us consider the energy scales involved. For most 
of the particles considered so far, the energy barrier height from the metastable 
state up to the saddle point is of the order of a few Kelvins whereas that from the stable 
state up to the saddle point is between $10^3$ and $10^5$ Kelvins. These energy scales 
should be compared with the energy necessary to kick the SQUID out of its metastable 
superconducting state which is of the order of few Kelvin. Therefore, only a small energy 
transfer is necessary to measure the magnetization reversal. In addition, a proper 
orientation of the easy axis of magnetization with respect to the SQUID loop can further 
reduce the coupling during the first stage of the magnetization reversal. In the case of an 
easy axis of magnetization perpendicular to the current direction in the microbridge, the 
coupling factor between SQUID loop and particle is proportional to (1 - cos $\varphi$), 
where $\varphi$ is the angle between the direction of magnetization and its easy axis, i.e. 
the coupling is very weak at the first stage of magnetization reversal.

The cold mode method allows one to detect the magnetization reversal some tenths of 
nanoseconds afterwards, i.e. the switching field can be detected very precisely. The main 
disadvantage of this method is that only the switching field of magnetization reversal can 
be measured and not the magnetization before and during the magnetization reversal. Note 
that the precise measurement of the time dependence of magnetization during the 
tunneling should be forbidden by Heisenberg's uncertainty relations.

Recently, we used an array of micro-SQUIDs for studying magnetic molecular clusters \cite{Science}. The sample is placed on top of the array of micro-SQUIDs 
so that some SQUIDs are directly under the sample, some SQUIDs are 
at the border of the sample and some SQUIDs are beside the sample. When a SQUID is very close to the sample, it is sensing 
locally the magnetization reversal whereas when the SQUID is far away, 
it is integrating over a 
bigger sample volume. The high sensitivity of this magnetometer allows us to study 
single molecular clusters crystals  of the order of 10 to 500 $\mu$m. 
The magnetometer works in the 
temperature range between 0.035 and 6~K and in fields up to 1.4~T with sweeping rates 
as high as 1~T/s, and a field stability better than a microtesla. The time resolution is about 
1~ms allowing short-time measurements. The field can be applied in any direction of the 
micro-SQUID plane with a precision much better than 0.1$^{\circ}$ by separately 
driving three orthogonal coils \cite{WW_PhD}.

\end{document}